\documentclass{article}
\usepackage[utf8]{inputenc}
\usepackage[round]{natbib}
\usepackage{paralist}
\usepackage{amsmath}
\usepackage{authblk}
\usepackage[normalem]{ulem}

\usepackage[margin=1.5in]{geometry}
\usepackage{graphicx}
\usepackage{rotate}
\usepackage{epsfig}
\usepackage{placeins}
\usepackage{dirtytalk}
\graphicspath{{pictures/}}
\newcommand{\ie}{\emph{i.e.}, }
\newcommand{\eg}{\emph{e.g.}, }
\newcommand{\avg}[1]{\langle #1 \rangle}
\newcommand{\Eloc}{E_{\text{loc}}}
\newcommand{\Eglob}{E_{\text{glob}}}
\usepackage[usenames,dvipsnames]{xcolor} 
\title{Modelling terrestrial route networks to understand inter-polity interactions (southern Etruria, 950-500 BC)}
\author[1,2]{Luce Prignano}
\author[1,2]{Ignacio Morer}
\author[3,4,5]{Francesca Fulminante}
\author[6,7]{Sergi Lozano}
\affil[1]{Departament de Fisica de la Mat\`eria Condensada, Universitat de Barcelona, Barcelona, Spain}
\affil[2]{Universitat de Barcelona Institute of Complex Systems (UBICS) Universitat de Barcelona, Barcelona, Spain}
\affil[3]{Department of Anthropology and Archaeology, Bristol University, Bristol, UK}
\affil[4]{Division of Humanities, Universit\`a degli Studi Roma Tre, Roma, Italy}
\affil[5]{Division of Archaeology, University of Cambridge, Cambridge, UK}
\affil[6]{Institut Catal\`a de Paleoecologia Humana i Evoluci\'o Social (IPHES), Tarragona, Spain}
\affil[7]{\`Area de Prehist\`oria, Universitat Rovira i Virgili, Tarragona, Spain}
\date{}
\begin{document}
\maketitle
\begin{abstract}
Ancient regional routes were vital for interactions between settlements and deeply influenced the development of past societies and their ``complexification''. At the same time, since any transportation infrastructure needs some level of inter-settlement cooperation to be established, they can also be regarded as an epiphenomenon of social interactions at the regional scale. 
Here, we propose to analyze ancient pathway networks to understand the organization of cities and villages located in a certain territory, attempting to clarify whether such organization existed and if so, how it functioned.
To address such a question, we chose a quantitative approach. Adopting network science as a general framework, by means of formal models, we try to identify how the collective effort that produced the terrestrial infrastructure was directed and organized.
We selected a paradigmatic case study: Iron Age southern Etruria, a very well-studied context, with detailed archaeological information about settlement patterns and an established tradition of studies on terrestrial transportation routes, perfectly suitable for testing new techniques.
The results of the modelling suggest that a balanced coordinated decision-making process was shaping the route network in Etruria, a scenario which correlates well with the picture elaborated by different scholars using a more traditional technique.
\end{abstract}

\section{Introduction}\label{sec:intro}
In the last few decades, there has been many attempts to identify the nature of the interplay between different human communities. The interaction between inhabitants of cities, towns, or villages has been analyzed both in contemporary contexts -- from the perspective of social sciences -- and in history. 

Around 40 years ago, Preston proposed one of the first quantitative approaches to the problem. He suggested that ``an appropriate methodology for the study of interaction should focus on movement of people, goods and capital, on social transaction and on the provision of service'' \citep{preston_1975}.

Already during the 1970s, but with an increasing intensity in the last ten years, great attention has been devoted to understanding settlement systems through the application of Central Place Theory, locational models, Proximal Point Analysis, as well as gravitation and other interaction models (see for example \citealp{ bevan_and_wilson_2013,nakoinz_2013a,davies_etal_2014,fulminante_2014,palmisano_2017,rivers2011network}, with references to previous studies). 

Such studies usually address settlement hierarchy in terms of relative importance of the sites, trying to figure out to what extent general factors (\eg topography or social-ecological advantages) can explain why some places become more prominent than others. Some of these approaches have been developed specifically to be applied to archaeological case-studies. The kind of data they take as inputs are limited to size and position of settlements, as that is frequently all that is available to archaeologists. 

The application of Preston's approach in archaeology to address questions beyond hierarchies and rankings is a less easy task, because of the heterogeneity and incompleteness of data on ``movement of people, goods and capital''. For example, although it is often possible to identify the origin and destination of an object, such data limitations make defining the intermediate steps really challenging. Still, in some cases, the archaeological record provides enough useful information on the exchange of goods or cultural influences among settlements (see for example \citealp{paliou_and_bevan_2016} on Bronze Age Crete, or \citealp{nakoinz_2013b} on Early Iron Age Germany or \citealp{da_vela_2014} on Archaic/Roman northern Etruria).

In this line, an even more difficult unresolved issue is whether and how settlements located in a certain territory were organized at the regional level.
The existence of a certain degree of regional organization can be tackled by techniques for the analysis of site distributions such as nearest-neighbor analysis (\citealp{PSG1979}). Hence it is possible to distinguish between randomly distributed and clustered (organized) settlements, whilst more complex matters remain out of reach.
To what extent did past communities cooperate or compete? Were they just struggling for their individual benefit or were they aware of their interdependence? These are just a few questions that can in principle be addressed quantitatively by means of formal models. But what kind of data is suitable to be used for hypothesis selection when dealing with this class of issues?

Here, we propose to gain a better understanding about these important topics by analyzing terrestrial transportation infrastructures (TTI). Indeed, the system of roads that existed in a territory might encode the footprint of processes and interactions at the regional scale. Starting from the Bronze Age, but even more with the advent of the Iron Age, the increasing social complexity and the accumulation of resources set the conditions for humans to have both the incentive and the capability to build roads \citep{earle_1991, lay_1992}. Constructed roads flourished along with the development of urban societies, when performing cuts, building bridges, or removing obstacles became both necessary and affordable. 

The importance of TTI for the understanding of the political and social organization of the communities that created and maintained them has been previously assessed for example in relation to the Roman Empire (\eg \citealp{chevallier_1976,taylor_1979,crumley_and_marquadt_1987,purcell_1990,mattingly_1997,malkin_2011}).  In the 1990’s  Trombold presented an important collection of works on TTI in the New World (\citealp{trombold_1991a}, followed by \citealp{jenkins_2001} and \citealp{smith_2005}). Recently, a renewed interest in TTI seems to have produced a number of new studies in Pre-Roman and Roman Europe (\eg \citealp{filet_2017,groenhuijzen_2016,matteazzi_2017}), suggesting that this is a growing field of research (see \eg recent projects such as ORBIS\footnote{http://orbis.stanford.edu/orbis2012/} by Stanford University, or the New Transhumance project in Toscana, \citealp{pizziolo_2016}).

Building on this literature, we propose to take a further step and try to infer aspects of the political organization of a region from the quantitative analysis of TTI. To this end, it is worth noting that not all kinds of roads will embed pertinent information. Paths had indeed different purposes, the most common ones being to connect: resources (\eg herds to pastures and water, or winter and summer pastures); humans and resources (\eg villages to pastures, water, mines); and human communities with each other (\ie settlements to settlements). Notice, that only ways of the last type were directly related with inter-settlement (villages, towns and cities) interactions. Consequently, we assume that roads of this kind were the output of a collective effort for the benefit of one or more of the parties involved. 

More specifically, we developed a baseline methodology to contrast hypotheses about the organization of a system of settlements, starting from a regional road map. Such a methodology consists of three fundamental ingredients: \begin{inparaenum}[(1)] \item a procedure for extracting relevant quantitative data from road maps; \item a set of competing hypotheses about organizational aspects of road construction; and \item formal models translating such hypotheses into mechanisms for generating synthetic data to be compared against the empirical ones \end{inparaenum}. 

The underlying idea is that some models reproduce relevant features of the empirical TTI with higher accuracy than others. Thus, we can determine which hypothesis (or hypotheses) better explains the empirical evidence and is therefore more likely to resemble the actual mechanisms of organization.  To develop such a methodology, we adopted network science as a general framework. We regard this as a natural choice, given that we chose to focus on road networks because of the information embedded in their connectivity and functionality. Network science provides us both an analytical toolbox for the characterization of such aspects of TTI and a conceptual framework for model building.

We selected a paradigmatic case study: Iron Age Southern Etruria, a very well-studied context, with detailed archaeological information about settlement patterns and an established tradition of studies on TTI (see Section  \ref{sec:background}). The political organization adopted by Etruscan cities is also known in its essential features - it was a league of loosely cooperating independent city-states (\textit{Duodecim Populi Etruriae}, ``The League of the Twelve people of Etruria'', on which see below). A number of scholars suggested studying roads and paths as a way of understanding the political and social organization (partially already \citealp{potter_1979}, p.79ff., \citealp{boitani_1985, izzet_2007}, and especially \citealp{tuppi_2014}). Therefore this makes the area an eminently suitable case-study for testing our new technique.

\section{Background and materials}\label{sec:background}
Between the beginning of the Iron Age and the Archaic Period, (southern) Etruria underwent a complex process of urbanization (see \eg \citealp{stoddart_and_spivey_1990,barker_and_rasmussen_1998,rasmussen_2005,bonghi_2005,pacciarelli_2001,pacciarelli_2010,pacciarelli_2017,riva_2010,marino_2015, stoddart_2017}). Contrary to what occurred for instance in the nearby region Latium Vetus, such a scenario did not lead to the emergence of any centralized authority or any noticeable disruptions in the balance between the city-states (on this see \eg \citealp{fulminante_and_stoddart_2012}). Even though the ways in which these Italian peoples were organized continues to be an enigma, it is well known that the Etruscan cities shared some religious values and that the Etruscan peoples would at times, possibly on an annual basis, congregate.

As referred by Livy, annual meetings (\textit{concilia}) of the Etruscan League were held at the \textit{Fanum Voltumnae}, the federal sanctuary dedicated to Voltumna and traditionally considered to be near Orvieto. They gathered to discuss political and military matters, such as the war between Veii and Rome around the end of the 5th and beginning of the 4th century BC. The mettings probably took place earlier as well, during the regal dynasty of the Tarquins (7th-6th century BC) or even Romulus (middle of the 8th century BC), or even in the preceding Early Iron Age.  

These sources on the Etruscan league suggest that in times of need, the cities probably did request military aid from one another, but there was probably little obligation to comply with such a request. Short-lived coalitions as well as smaller alliances, especially in the southern territory, would likely have been formed for defense and commercial purposes, and perhaps some civic leaders met to discuss political and military situations (on the Etruscan League and the political life of Etruscan cities see the useful reconsideration by \citealp{gillett_2010} and recently \citealp{tagliamonte_2017}). 
The status of larger cities was different from that of other settlements over whom they exerted significant control.  The location of smaller secondary sites and even rural sites within the landscape have been shown to be integrated into the definition of the territorial boundaries of the city (see \eg \citealp{de_santis_1997} on the territory of Veii and \citealp{damiani_and_pacciarelli_2007} on the territory between Veii, Caere and Rome), and the evidence of roads and other engineering works, agricultural practices and settlement patterns has also shown that the rural landscape was increasingly differentiated from the urban landscape (\citealp{izzet_2007}, p.193ff. with previous references).

Summarizing, the Etruscan cities controlled a territory that included other minor towns and villages and occasionally organized with each other making some coordinated decisions, cooperating for some aspects, and competing for others. In this complex landscape of interactions, the establishment and maintenance of a road network was of crucial importance for the functioning of Etruscan polities both at local and regional scales. Roads linked the major centres to each other, to minor centers and even to rural sites. 

Such a network was the result of a significant collective effort, at least since the Etruscans started to build heavily worked or manufactured roads. Road-building is attested in different parts of Etruria, though most notably in central southern Etruria, and has been largely studied. 
As observed by Erik Wetter, the principal roads of antiquity were drawn \say{in quite informative and reasonably accurate fashion} already on \say{an interesting black-and-white map of the \say{Territorio o Distretto di Roma} published in 1674 by Geografo Pontificio Innocenzo Mattei, who based it on studies started in 1628 by the German scholar Lucas Holstenius} (\citealp{wetter_1962}, p.172). Later, Roman roads and Etruscan paths have been the object of study by topographers such as Nibby, Gell, Ashby, Nissen, Solaris, Lanciani and finally Lugli (who founded the Roman School of Topography) and/or landscape archaeologists such as Potter and Ward-Perkins (see \citealp{ward_perkins_1962a, ward_perkins_1962b} and discussion in \citealp{potter_1979}). This literature is well summarized and presented in the synthesis by Vedia Izzet (\citealp{izzet_2007}, pp. 193–207).

The dating of roads is a difficult task that can be done indirectly, based on that of settlements they pass between and the objects from the tombs that lined them or from the material from settlements associated with them (\citealp{potter_1979}, p. 80ff. but similar observations in \citealp{trombold_1991b}). Nonetheless, Timothy Potter, after J. Ward-Perkins, has differentiated between the winding, irregular courses of the Bronze Age and Villanovan paths and more recent longer roads that were more suited to wheeled transport (\citealp{potter_1979}, pp. 79–84). The chronology considered in the present work (from the end of the Bronze Age to the end of the sixth century BC) corresponds to the period of progressive replacement of early mule and pedestrian routes by what Potter termed ‘engineered’ roads, suitable for wheeled transport which prevailed during the seventh and sixth centuries\footnote{Potter connects this transition with the intensification of trade and the diffusion of the wheeled cart during the Orientalizing Period (\citealp{potter_1979}, p. 81). As mentioned earlier while this evolutionary perspective is still valid, recent work by Ulla Rajala on the TTI in the territory of Nepi (\citealp{rajala_2002}, chapter 6.4) and Miriam Rothenberg on the TTI around Veii has demonstrated that the relationship between pre-and proto-urban paths and urban roads might be more complex \citep{rothenberg_2014}.}.

Many of the cross-country Etruscan roads were not based on natural corridors in the landscape in the same way as the earlier tracks. Instead, they followed longer routes that exploited gentler gradients, and the latter were minimized by cuttings which vary in shape and scale (\citealp{izzet_2007}, p. 194). Recently the road-cuts of Etruria have been reconsidered and mapped in an important work by Juha Tuppi, who also emphasized their role in the creation and maintenance of urban authority over the territory \citep{tuppi_2014}. Additionally, roads were frequently complemented by complex engineering works (fords, bridges, ditches, culverts, etc.).

All these significant engineering and hydraulic works implied major manipulations of the landscape. Consequently, Etruscan roads can be understood as the product of an organized collective effort by the Etruscan cities. This supports the 'reverse engineering' approach proposed here, based on inferring how Etruscan cities did organize by analyzing the structure of their road network.

To pursue this goal, we relied on the vast and rich literature available. The landscape of Southern Etruria during our period of study has been intensively analyzed, and relevant efforts have been devoted to map both the main settlements and routes connecting them. To identify the main settlements and their location, we relied mainly on the Repertorio dei Siti Preistorici e Protostorici della Regione Lazio \citep{belardelli_etal_2007} and the study by Iaia and Mandolesi \citep{iaia_and_mandolesi_2010} for the Early Iron Age, and on the work by Marco Rendeli on the territorial organization of southern Etruria for the Orientalizing and Archaic Ages \citep{rendeli_1993}.

\begin{figure}[ht]
\centering
\includegraphics[width=0.8\columnwidth]{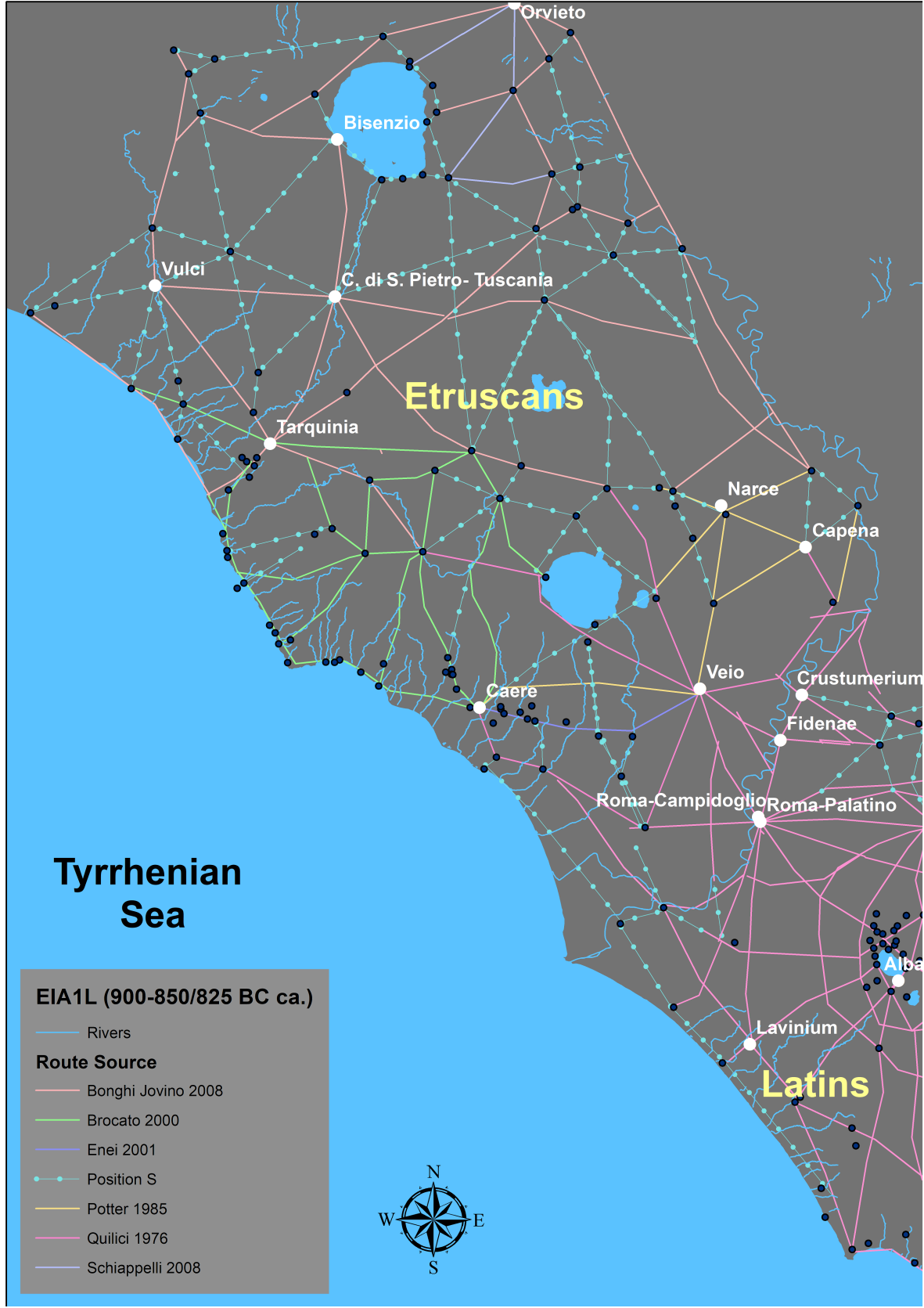}
\caption{Schematic Map of the Etruscan routes. Early Iron Age 1 Early (950/925 – 900 BC). The routes labelled ``S Position’’ are routes hypothesized on the basis of the position of secondary settlements.}
\label{fig:fig1}
\end{figure}

These sets of settlements were updated according to the volumes of Studi Etruschi, together with the series of conferences and exhibition catalogs or publications that followed them (for example \citealp{della_fina_2013}). Concerning routes, we compiled those hypothesized by several authors \citep{potter_1979,zifferero_1995,tartara_1999,brocato_2000,enei_2001,bonghi_2008,schiappelli_2008,wetter_1962}. Figure~\ref{fig:fig1} shows the hypothesized Etruscan terrestrial communication routes in the Early Iron Age 1 Early (950/925 – 900 BC) according to the various authors proposals. We also included routes hypothesized on the basis of the position of secondary settlements, especially if they were present on the alignment of a later Roman road. This approach is well known and established in topographic studies both in Italy and in the new world (see \citealp{trombold_1991b}).
\FloatBarrier
\section{Etruscan road networks}\label{sec:charac}
Adopting a network science approach implies that the first step we have to take is to translate available information on pathways from the usual map format into networks, \ie mathematical structures made up of interconnected objects. Once the empirical system is mapped onto weighted geographical networks, one can apply the established analytic tools provided by network science for their characterization.
\subsection{From maps to networks}\label{sec:map2net}

The task of translating road maps into networks is not straightforward and can be performed in many alternative, not equivalent ways. It is important to proceed carefully since decisions made at this stage will condition future results and their interpretation at a very general level. We need a definition of the basic formal ingredients that must be consistent with the questions that we want to address. Such ingredients are: a set of objects, called nodes or vertices, and a set of connections, called links or edges. 

The usual representation of road networks takes each cross-point as a node (\citealp{barthelemy_2011}). In this way, each connection between two crosses can be naturally translated into a link, thus obtaining a planar graph, \ie a graph where edges do not cross. This simple and elegant construction is perfectly suitable for studying traffic problems (\eg congestion) and communication properties such as the navigability of the network. 

Nonetheless, our goals are different and this is not the best approach for our purposes. In particular, since we are studying inter-settlement interactions, we need our nodes to be human communities with a certain degree of political agency, such that they could play an active role in shaping the regional infrastructure. Consequently, links also need a new definition. The simplest option is to consider that a bidirectional link between two sites is established whenever they were directly connected by a terrestrial route, with no other settlement in between.
Once the rules are set, the second stage consists in selecting and organizing the empirical data. 

Since during the chronology under study southern Etruria underwent deep transformation, we decided to divide the period of interest into different temporal stamps.
We determined the largest possible time spans during which a set of settlements coexisted without any major changes. Thus, we obtained five time stamps, each one with its corresponding list of geo-localized sites:
\begin{itemize}
    \item Early Iron Age 1 Early (EIA1E): 950/925 – 900 BC
    \item Early Iron Age 1 Late (EIA1L): 900 – 850/825 BC
    \item Early Iron Age 2 (EIA2): 850/825 – 730/720 BC
    \item Orientalizing Age (OA): 730/720 – 580 BC
    \item Archaic Period (AA): 580 – 500 BC
\end{itemize}
We then resolved whether and how those sites were connected, that is, if they were directly linked by a terrestrial route. We assumed that, since within each one of the five Ages considered, the set of settlements can be regarded as constant, the same applies to the routes connecting them. In this way, the subject of the analysis is reduced from a continuously evolving ensemble of settlements and routes to five static networks.

Finally, since we were interested in terrestrial routes as the product of a collective effort, requiring the allocation of resources to be built and maintained, it was essential to somehow quantify their cost. For modelling purposes, we could not settle for assigning a cost only to the links that are present in the empirical networks. What we needed was a mathematical function able to assign a cost to any potential links, that is, to any connection between any pair of settlements. Thus we are able to compare empirical and model-generated networks. 

\begin{figure}[ht]
\centering
\includegraphics[width=0.6\columnwidth]{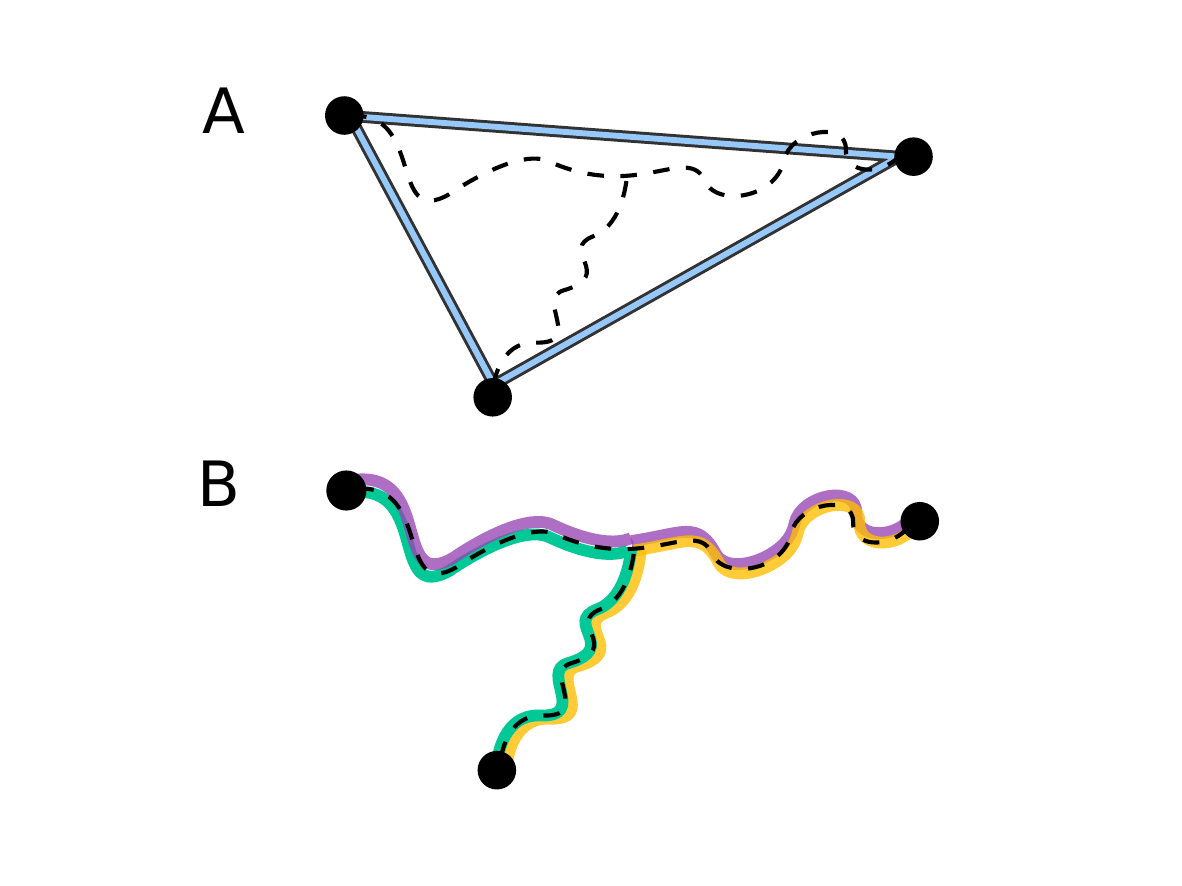}
\caption{Example of three hypothetical settlements connected by paths that share stretches. In the upper sketch (A), the cost of each connection is approximated by the geodesic distance; in the lower one (B), they are estimated through the walk between them. In this way, each stretch is counted twice.}
\label{fig:fig0}
\end{figure}

So the best option was to rely on geographical information, discarding detailed data about construction (\eg road width) which is only available for some roads, but not for others. It is indeed reasonable to assume that, beyond the peculiarities of road building in each individual case, the cost of a road is roughly proportional to its length. To determine the length associated to each connection, we could have implemented GIS based analysis, measuring it directly in the case of known ways and adopting a least-cost path (LCP) approach for those paths whose route is not completely known or those that were not part of the empirical networks. However, using different levels of precision for different links might be detrimental. Alternatively, in principle, it would have been possible to calculate the LCP between every pair of connected nodes, regardless what we know or do not know about their roads. Nonetheless, doing so without taking into account the existence of common stretches (see Figure~\ref{fig:fig0}) between different paths whose length contribute to the cost of more than one connection cannot be considered an improvement with respect to much simpler options, such as straight line distances. It rather constitutes a time consuming alternative with doubtful advantages. On the other hand, to explicitly account for stretches that belong to two or more paths is not just a very complicated task, but would have also significantly affected the characterization and modelling of such networks, that is, networks whose links have costs that depends of the existence of other links.
We concluded that the optimal way to address the geographical factor was to take into account only essential aspects, while a precise method to measure road lengths was ruled out.
We represented sites as geo-localized nodes and assigned weights to the links according to the geodesic distance between the nodes they connect. This was a quite good approximation provided that the region under study was relatively small and homogeneous, but more importantly, the lack of precision is evenly distributed among the nodes, without biases towards less studied areas. Additionally, it matched well the scope of our methodology, which focuses on the general properties of the system, not on the details of individual paths.
\subsection{Characterization of terrestrial route networks}
Usually, archaeological studies have looked at the local scale to hypothesize individual routes, taking into account local evidence and geographical or environmental considerations to propose connections. On the contrary, in the present work, we were interested at looking to such connections altogether, that is, to the structure that they formed at the regional scale. By doing so, we could see a transportation network with particular features. These features conditioned the way the TTI functioned, in the sense that they determined its performance, \ie the efficiency and robustness of the communication that took place on it. Such systemic features are not defined by individual connections between specific pairs of settlements, but by the whole of them. For instance, we might overly focus on the presence of central places much better connected than many peripheral ones (inequality) or the existence of routes or settlements that, if inaccessible, made the network fall apart (fragility).

The very first step for a proper characterization of these systems was the description of the overall connectivity. Here, the spatial nature of the transportation network required not only the consideration of the amount of connections but also their length. The strength $s_i$ of a node $i$ (also known as weighted degree) extends the idea of node degree $k$ (number of connections) to the weighted case. Given a node $i$, $s_i$ measures the total length of its adjacent links (\citealp{barrat_et_al-2004}):
\begin{equation} \label{eq:strength}
s_i = \sum_{j\in V} l_{ij}
\end{equation}
where $V$ is the set of neighbors of $i$, and $l_{ij}$ is the length of the link between $i$ and its neighbor $j$, that is, the distance between the two nodes. The average strength $\avg {s}$ is the mean value over the set of nodes:
\begin{equation} \label{eq:avgkw}
\avg {s} = \frac{1}{N} \sum_{i=1}^{N} s_i
\end{equation}
Likewise, it was informative to know the mean length of all the links. For example, considering the same amount of total length in a system, it could be distributed in lots of short edges or in very few long ones, which implied quite different values of the mean length. The average edge length $\avg {l}$ is obtained by simply averaging the distance of all the links in the system:
\begin{equation} \label{eq:avgl}
\avg {l} = \frac{1}{M} \sum_{i=1}^{N} \sum_{j>i\in V} l_{ij},
\end{equation}
where ? is the total number of links in the network. Notice that these first two indices also provide the average degree $\avg k$, which is obtained by dividing $\avg {s}$ by $\avg {l}$.

Another classical property in network analysis, the clustering coefficient, describes the tendency of nodes to form dense groups (clusters). Given a node $i$, $C_i$ indicates the proportion of all potential links between the neighbors of $i$ that actually exist:
\begin{equation} \label{eq:clust_i}
C_i = \frac{m_i}{k_i(k_i-1)/2}
\end{equation}
where $m_i$ is the number of links connecting two neighbors of $i$, thus closing a triangle containing $i$. Averaging this ratio over the set of nodes, the average clustering coefficient $\avg C$ constitute an indicator of the density of closed triangles in the network:
\begin{equation} \label{eq:clust}
\avg C =  \frac{1}{N} \sum_{i=1}^{N} C_i
\end{equation}
A high (or at least, not negligible) clustering is commonly found in many real world networks \citep{albert_2002}. The main point of calculating $\avg C$ is to indicate the probability that two neighbors of a site are also connected to each other. Hence, there is no need to use other definitions of clustering adapted to weighted networks. In the specific case of route networks, such coefficient indicates how often two settlements ``choose'' to build a direct connection even if they are already connected through a common neighbor.

Besides these generic metrics, we needed specific measures able to capture the nature and functionality of these particular networks. It is a well known fact that the topology of a network is closely related to its functionality: from the brain to the Internet, from trophic relations to metabolic networks, the way all these systems work is closely related to the organization of the connections \citep{boccaletti_2006}. TTI are no exception. They enable interactions between sites, a process that ultimately comes down to some sort of transfer (information, goods, etc.) among sites across a geographic space. Building from this idea, in the past fifteen years, different measures related to the concept of efficiency of communication in networks embedded in a physical space have been introduced.
A first type of measures addresses the efficiency of a network on a global scale. These quantify how well the information is exchanged across the whole network, assuming that the closer two sites are, the easier information is transferred between them. In its simplest definition, the efficiency of the communication between two sites is calculated as the inverse of the shortest path length -- that is the minimum number of links -- separating them \citep{latora_2001}. However, dealing with geographic networks, we needed a definition that compares the best (ideal) path to the shortest path provided by the network. The shortest weighted path length $L_{ij}$ between two sites $i$ and $j$ is the path with the lowest sum of distances, no matter the number of links in it. As a consequence, a path of many steps could be shorter than a path of just few steps if the sum of their lengths was smaller\footnote{It is worth noting that, since we were not taking into account the number of nodes in the path, the implicit underlying hypothesis was that it was possible to travel through any settlement in the region without any additional cost or delay.}.

Following \citep{vragovic_2005}, we considered the geographic distance divided by the (weighted) shortest path length as an appropriate measure for the efficiency in the communication between two sites. The global efficiency of the network, $\Eglob $, is calculated as the average over all pairs of nodes:
\begin{equation} \label{eq:eglob}
\Eglob = \frac{1}{N(N-1)} \sum_{i\neq j} \frac{d_{ij}}{L_{ij}}
\end{equation}
According to this definition, the global efficiency $\Eglob$ of a fully connected network -- where each and every pair of nodes are connected -- is equal to 1, no matter the number of nodes, neither their positions.

On the other hand, the local efficiency of a node $i$,  $\Eloc (i)$, was proposed to quantify how well information is exchanged between its neighbors when that particular element is removed\footnote{Even though local efficiency involves the study of a neighborhood, it is not exclusively dependent on clustering coefficient, that is, the links between one’s neighbors. Therefore the two measures can be regarded as complementary.}. This metric estimates how resilient a network is against localized failures. We adopted the definition in \citep{vragovic_2005}, devised specifically for geographic networks. Mathematically,
\begin{equation} \label{eq:eloc}
\Eloc (i) = \frac{1}{k_i(k_i-1)} \sum_{j\neq k \in V} \frac{d_{jk/i}}{L_{jk}}
\end{equation}
where $d_{jk/i}$ is the path between $j$ and $k$, both neighbors of $i$, if we would remove $i$ from the system. Averaging over all the nodes, a measure of the local efficiency ($\Eloc$) of the network under consideration is obtained.

After exploring other network properties, we concluded that these five (\ie $\avg {s}$, $\avg {l}$, $\avg C$, $\Eglob$ and $\Eloc$) are adequate and sufficient to characterize the system. The first three provided enough information about the overall connectivity (number and length of the connections, fraction of closed triangles), and the efficiency measures indicate to what extent the topology was suitable for certain dynamic processes (\eg information exchange) taking place on the network. After trying several network generator models, \eg random geometric graph or spatial Erdos-Renyi graph (see~\citealp{barthelemy_2011}) with the same node positions and total link length, we concluded that it is not possible to obtain two networks with very similar values of these five measures that, at the same time, have very different values of other common structural indexes (\eg assortativity, degree centralization, strength centralization, average shortest path length \citep{albert_2002,boccaletti_2006}).

Tab.~\ref{tab:emp} provides a characterization of the empirical networks from these 5 structural features. Among all the considered metrics, whose implications are discussed in further details in the Section \ref{sec:validation}, we would like to highlight the high global efficiency of these networks, despite the relatively low connectivity (the corresponding fully connected networks have a total amount of connection length that is between $150$ and $300$ times that of the empirical ones). Hence, we had a preliminary indication of the good quality of the design of Etruscan terrestrial infrastructure.
\begin{table}
    \centering
    \begin{tabular}{ | l | r | r | r | r | r |}
      \hline
      Age & EIA1E & EIA1L & EIA2 & OA & AA \\ \hline \hline
      $N$ & 116 & 115 & 130 & 168 & 179 \\ \hline
      $\avg {s}$ (km) & 27.35 & 28.88 & 27.88 & 24.84 & 26.00 \\ \hline
      $\avg {l}$ (km) & 7.792 & 8.021 & 7.711 & 6.631 & 6.287 \\ \hline
      $\avg C$ & 0.143 & 0.184 & 0.173 & 0.236 & 0.270 \\ \hline
      $\Eglob$ & 0.875 & 0.887 & 0.869 & 0.872 & 0.888 \\ \hline
      $\Eloc$ & 0.507 & 0.571 & 0.606 & 0.657 & 0.722 \\ \hline
    \end{tabular}
    \caption{Network properties of the empirical systems. For each period we report the system size $N$, the average strength $\avg {s}$, the average edge length $\avg {l}$, the global efficiency $\Eglob$ and the local efficiency $\Eloc$.}
    \label{tab:emp}
\end{table}
\section{Network modeling}\label{sec:model}
In the present study, we are developing a methodology to infer how settlements were organized at the regional level by analyzing the structure formed by the roads that connected them. However, such a methodology cannot consist of a mere analysis of network properties since it would never be conclusive, unless we were able to link such properties to the mechanisms that generated them. Therefore, to take a step further, we had to figure out if there was some general rule governing the decision-making processes that shaped the terrestrial route network. It was necessary to hypothesize generative mechanisms that might had created the empirical network and to contrast their different outcomes (artificial networks) against the empirical evidence. In other words, we had to devise competing network models, each one corresponding to a strategy according to which the nodes made decisions about which links had to be established.

Such a decision-making process needed to be conceptualized in a simplified way to make them possible to be reformulated as a mechanism for the creation of connections. We wanted to keep our models as simple as possible to have clearly distinguishable options, directly related to different hypotheses that could be contrasted against the empirical evidence. Simplicity is a fundamental ingredient not just for immediate practical purposes, but also for the sake of generality. We were looking for models that can applied to a variety of analogous case-studies that differ from the present context in everything but some very fundamental aspects of the interaction between settlements. By building complicated models based on the details of the present case-study, we would have made future applications more difficult.
In this regard, let us recall that we aimed to reproduce patterns and quantitative properties of the empirical networks, features clearly related with how decisions were made, disregarding local details --  \ie the presence or absence of individual connections -- which are likely due to mere contingency.

Since we wanted to focus on the decisions behind the creation of links, we had to set up our experiments (simulations) in such way that those decisions were made under similar conditions compared to the empirical networks. If we were attempting to disclose how the collective effort that produced the terrestrial infrastructure was directed and organized, then we could not modify the general situation in which such an effort took place. In particular, we considered that the set of settlements with their corresponding geographic locations and the amount of available resources -- here quantified as the total link length -- had to be the inputs of the models, not the parameters to be fitted. We were not trying to understand why in southern Etruria there were not more or less or different settlements, or more or less roads. We are addressing the question of why and how the settlements that existed in the region built those roads instead of others.

This approach has implications that we would like to discuss. Firstly, we were considering the economic development of the region, which was supposed to fix the costs the region could afford in terms of kilometers of routes, as a process unrelated with the TTI. Although this is obviously not true -- roads were vital to the development and there was a clear feedback loop -- there were so many complex factors at play that it is not feasible to model such relations in a simple formal way. Therefore, to assume that the cost of the network at a given time is fixed externally, was the most reasonable (and useful) assumption at hand. 

Secondly, we have to clarify that we were proceeding in this way for each Age independently. Hence, we were considering the existence of links in a given Age without considering if they existed in previous times. If a certain road existed during a certain Age, it was because it had a purpose at that time, without regard to what may had happened before. The underlying assumption is that the effort for track maintenance was as important as the cost for creating routes anew and that no maintenance was the same as destruction. Although it is not easy to corroborate this hypothesis, it is reasonable to assume that, over several decades (\ie the length of our time-stamps), the efforts  necessary for the maintenance equaled those of the construction. Indeed, some special cases of which we have direct archaeological evidence, provide good indication of this. The \textit{Vie Cave} - roads that needed to be recut over and over to smooth their worn surface, to the point that they became actual trenches (\citealp{izzet_2007}) - suggest that maintenance was as expensive as road building, at least, in the long term.

This conceptualization set the basis for the design of a quantitative study based on network analysis and network modeling. In the following sections, we introduce three very simple models without parameters. Each one emulates a different scenario in terms of \begin{inparaenum}[(a)] \item the information available to each settlement and \item the nature of the relations between them \end{inparaenum}. Since there was no centralized authority ruling the region, the three of them assumed that settlements defended local interests and no scenario of strategic planning at the regional level was included.
\subsection{Model L-L. Local information, Local decision making.}
The first model was inspired by the simplest hypothesis of all, so simple that it could be regarded as a ``null-hypothesis'': Sites individually competed to gain the largest portion of resources (\ie connection length).
The basic idea is that any settlement that had the opportunity to collect the means to build a new connection, did it, thus decreasing the chances of doing the same for other settlements. Since we were not interested in modelling the details of such a competitive process -- rather, we wanted to study the effect of competition as an abstract mechanism -- we let  chance determine who won and who lost.
Random selection of the nodes that are going to create a new connection is the most simple way to simulate the output of a competition whose dynamics are unknown. Thus, nodes that were more successful -- that is, those that were selected more often than others -- were able to gather a larger amount of resources.

Concerning the choice of the new connection that a selected node decided to build, we imagined that nodes had only local information, \ie they knew where the closest settlements were located and whether they were directly connected to them.
Therefore, in formal terms, model L-L takes two steps to establish a link:
\begin{enumerate}
    \item A node $i$ is drawn at random from the total population.
    \item A link between $i$ and its closest neighbor $j^*$ (not already connected to $i$) is created, regardless whether there was an already existing path connecting them through other settlements. 
\end{enumerate}
We start with a set of isolated nodes (no links at all), then steps 1 and 2 are repeated until the sum of all the link lengths reached that of the empirical system.
\footnote{Interestingly, our model L-L can be seen as a stochastic version of the networks generated for proximal point analysis, in the sense that in both cases nodes create links with those that are geographically near. The main difference is that in the case of model L-L the amount of neighbors that can be connected is not fixed, but fluctuates from node to node.}

Running this model many times, we created a huge number of networks.
Potentially, we generated all the networks compatible with our hypothesis of competing settlements provided only with local information. 
\subsection{Model G-L. Global information, Local decision making.}
In this second model, nodes had information about direct connections to local neighbors (as in model L-L) as well as indirect connections to other nodes (\ie paths including one or more intermediate nodes). The procedure to create links mimicked that of model L-L, but the criterion considered in step 2 changed in order to capture the additional information available to settlements. Hence, instead of always choosing the nearest geographical neighbor, the model takes into account a normalized distance $R$ that is calculated as follows:
\begin{equation} \label{eq:detour}
R_i(j) = R_j(i) = \frac{d_{ij}}{L_{ij}}
\end{equation}
Where $d_{ij}$ is the geographic distance between node $i$ and node $j$, and $L_{ij}$ is the length of the shortest existing path between them. Accordingly, model G-L followed these steps:
\begin{enumerate}
    \item A node $i$ was drawn at random from the total population.
    \item $R_i(j)$ values were computed for all nodes $j$ in the system (except for $i$).
    \item The node $j^*$ with the minimum $R_i$ was selected and a link between $i$ and that node $j^*$ was created.    
\end{enumerate}
The function $R_i(j)$ balances costs and benefits, prioritizing those links that shorten long paths (large $L_{ij}$) while wasting little resources (short $d_{ij}$). As in model L-L, steps 1, 2, and 3 were repeated until the sum of all the link lengths reached that of the empirical system being modeled.
Notice that, in principle, nodes provided with more information than in model L-L should be able to make 'clever' decisions limiting the waste of resources (\ie creating roads making non significant contributions).

Once again the model had a stochastic component, that is, it was partially ruled by chance. Therefore, each realization had its own microscopic characteristics, but all of them shared some macroscopic commonalities.
\subsection{Model EE. The Equitable Efficiency Model.}
This third model not only provided nodes with global information about their connectivity, but also with the ability to make coordinated decisions. More concretely, each settlement knows the length of each one of the existing paths between its location and any other settlements, as in model G-L, but this time they prioritized links globally according to their $R_i(j)$ value. The exact procedure of model EE is the following:
\begin{enumerate}
    \item For each node $i$, all the $R_i(j)$ values were calculated. 
    \item Each node $i$ proposed the creation of a link between itself and a node $j^*$ such that the $R_i(j^*)$ was the minimum value among all the $R_i(j)$ (local interest expressed by node $i$).
    \item All the proposals were ranked according to their $R$ value and a link was created between the pair corresponding to the global minimum (coordinated decision-making).
\end{enumerate}
Step 1, 2, and 3 were repeated until the summed lengths of all created links reached that of the empirical system.
The existence of a global priority list instead of individual, local ones emulated a certain degree of coordination among the settlements in the sense of a general intention of preventing weak-points in the network. In this way, without altering the nature of the interests of each settlement that were still local, we introduced a balancing mechanism to smooth out the effect of competition.

Even though this model did not explicitly optimize any global metric, and therefore could not be regarded as a model of global planning for the infrastructure, the coordinated decision making made it different from the previous ones in a very fundamental aspect. As explained above, L-L and G-L models were stochastic because the focal node $i$ was chosen at random. This was not the case in the EE model, since links were established between the pairs of nodes $i$ and $j$ that minimize $R(i,j)$ at each step. Chance did not play any role in this algorithm, which implies that the model’s final output was unequivocally determined: for each empirical network, model EE generated a unique artificial counterpart.

\begin{figure}[h!]
\centering
\includegraphics[width=0.7\columnwidth]{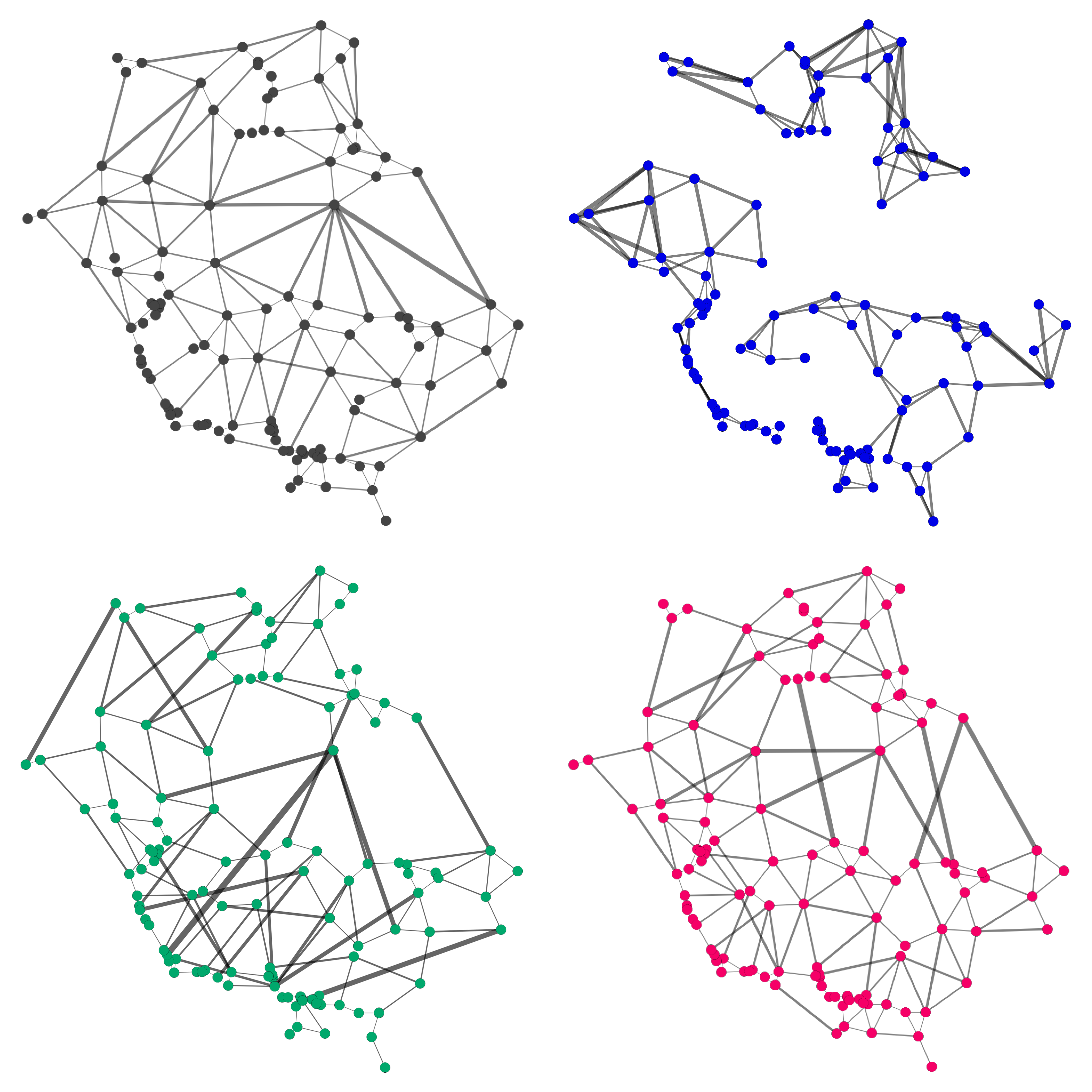}
\caption{Example of network realizations: Networks of Age EIA1L. Top left: empirical; top right: a random characteristic realization of model L-L; 
bottom left: a random characteristic realization of model G-L; bottom right: model EE.}
\label{fig:examples}
\end{figure}

As an illustrative example of the kind of connection patterns generated by the three models, in Figure~\ref{fig:examples} we represented the empirical network of Age EIA1L with its artificial counterparts: one realization of models L-L and G-L and the output of the EE model.
\FloatBarrier
\section{Validation of network models}\label{sec:validation}
In order to validate the three proposed models, we tested whether the synthetic networks that they generated were capable of reproducing the structural features (\ie $\avg {l}$, $\avg C$, $\Eglob$ and $\Eloc$\footnote{We are not including the average strength because, by construction, it takes the same value as in the empirical networks for all the model generated counterparts since $\avg{s} = \sum_{i=1}^N \sum_{j \in V_i}l_{ij} / N = 2 L_{tot} / N$, where $L_{tot}$ is the total link length.}) of the five empirical networks. As discussed above, we were not aiming to reproduce each and all of the connections. Our goal was to explain those features of the overall topology previously selected for their relevance in determining the performance of a transportation network.
This approach has an additional advantage when validating the models using empirical networks: It is not dependent on local individual details of the empirical networks. As already explained, terrestrial routes were inferred from a vast and heterogeneous amount of information. While first order connections among main primary centers (cities) are more certain, local connections among small villages might have slightly different interpretations. This uncertainty undermines the utility of validation by comparing individual connection one by one. On the contrary, average properties of the connectivity pattern would not change noticeably if a few of the routes were actually different from the ones considered here. Indeed, since our approach relies on system-scale information, we did not need the route map to be perfectly accurate for this methodology to be applied. 
\subsection{Network properties}
\begin{figure}
\centering
\includegraphics[width=\columnwidth]{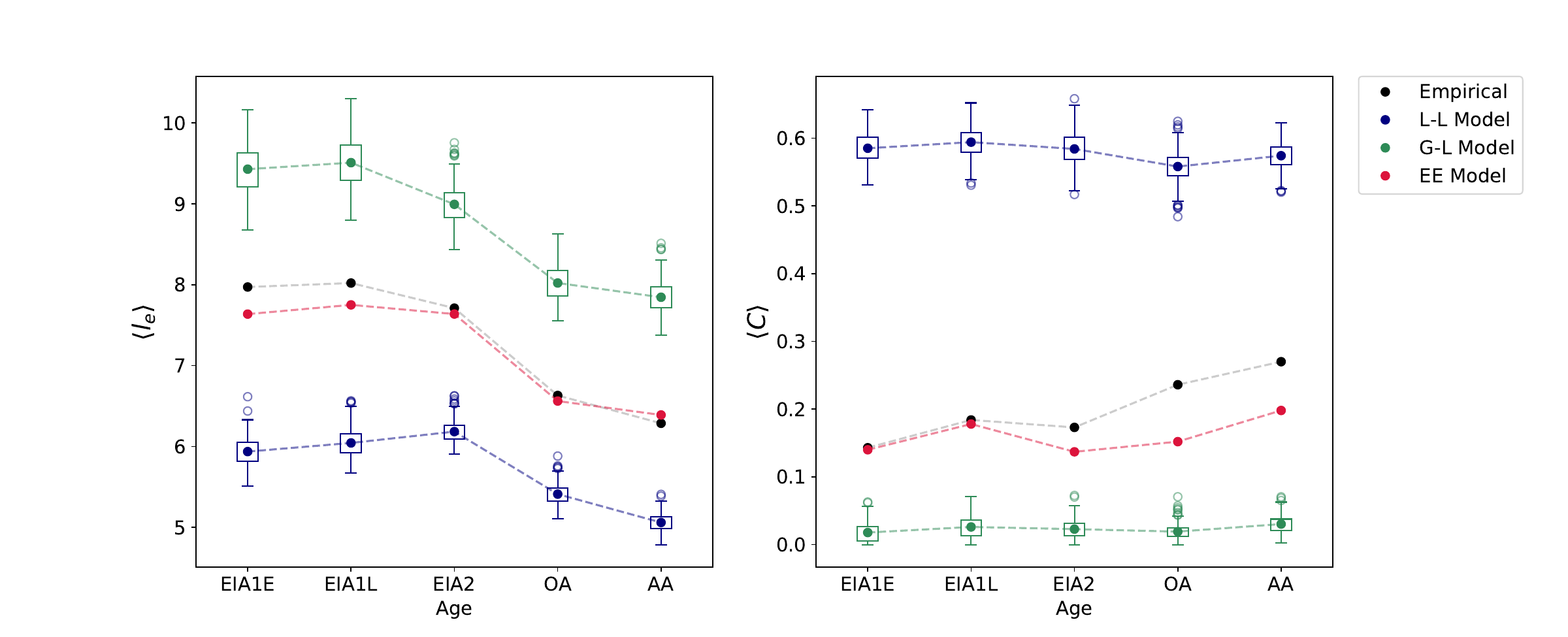}
\caption{Values of $\avg {l}$ (left panel) and $\avg C$ (right panel) for the empirical and synthetic networks. Points stand for average values, boxes span from the lower to the upper quartile values, and whiskers show the range of data.}
\label{fig:avg_l_clust}
\end{figure}

\begin{figure}
\centering
\includegraphics[width=\columnwidth]{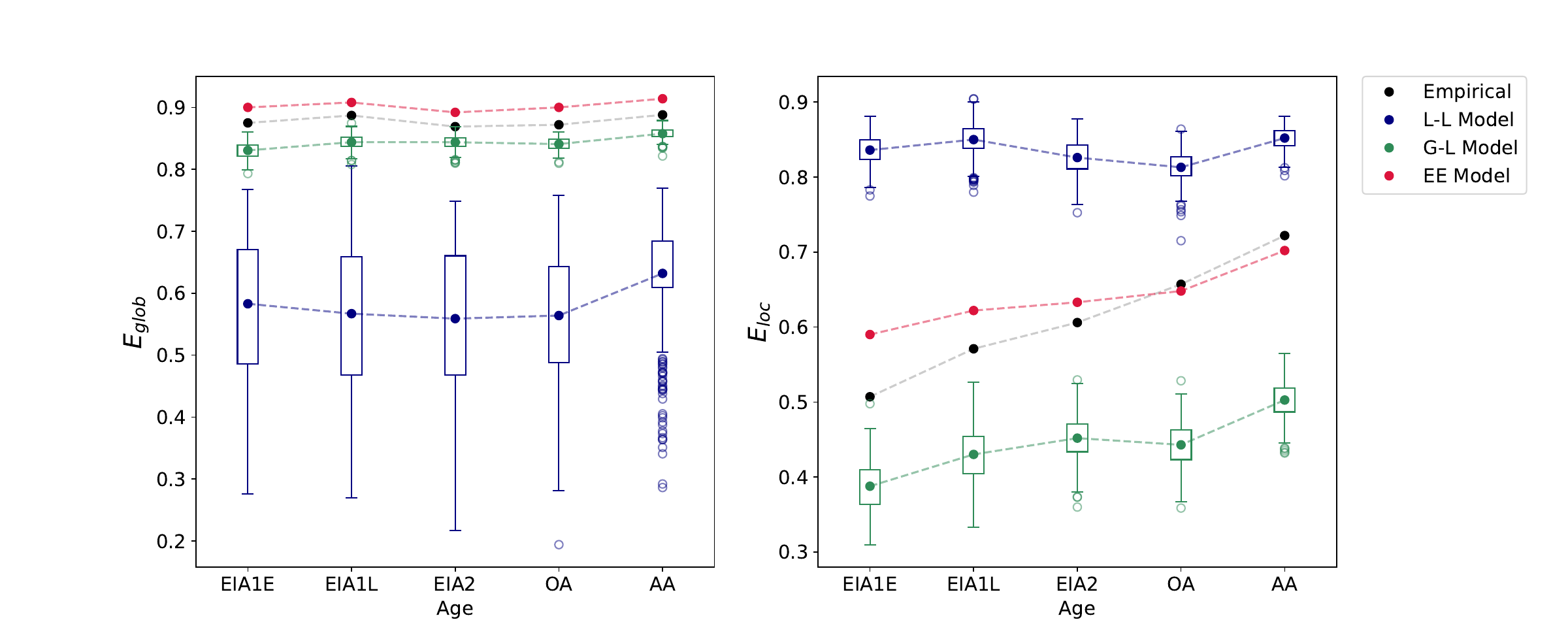}
\caption{Values of $\Eglob$ and $\Eloc$ for the empirical and synthetic networks. See Figure ~\ref{fig:avg_l_clust} for more details.}
\label{fig:effs}
\end{figure}

Figures~\ref{fig:avg_l_clust} and \ref{fig:effs} show, separately, the comparison of the properties calculated for the empirical networks and the output of the models along the five Ages. Given that model L-L and G-L are stochastic, we had to average over several realizations (repetitions of the modelling process) to obtain representative outputs. The number of realizations for each model and Age was not fixed \textit{a priori}. We kept generating new ones until all the average values of all the considered metrics and their standard deviations stabilized within an error of 1\%.

While significant differences with the empirical networks were observed in the cases of model L-L and model G-L, the EE model was able to recreate with good accuracy the relevant features for all five considered periods. In particular:
\begin{enumerate}
    \item When comparing with the empirical networks, the average edge length $\avg {l}$ was generally lower for networks produced through model L-L and higher for networks produced by model G-L. This is due to the fact that nodes in model L-L tended to connect with those that were geographically close to them. On the contrary, in model G-L, if drawn more than once, they looked for shortcuts towards nodes that lie at greater distances. As for model EE, the obtained values were intermediate and close to those of the empirical networks, the largest difference being found in the first period (up to $4.2\%$ with respect to the empirical value). In the later ages, this deviation decreased, becoming nearly imperceptible in the last three. In model EE, nodes tried to create shortcuts as in model G-L, but the prioritization mechanism forced the construction of the shortest ones first.
    \item The average cluster coefficient $\avg C$ was not reproduced correctly by either L-L or G-L models. With the L-L model we obtained values that were (on average) two or three times the empirical corresponding ones. This difference was higher for the first period (EIA1E), but it progressively diminished for later periods, as the value of $\avg C$ in the empirical system tended to increase through time while in the artificial networks it is kept constant. It is worth noting that, since two nodes close to a third one are also close to each other, model L-L naturally created many triangles, that is, networks with a very high clustering coefficient. As for model G-L, since the combination of random node selection and shortcuts were very unlikely to close triangle, its synthetic networks presented negligible values of clustering coefficient. Again, model EE provided the best results. It was the only outcome that reached values close to empirical ones (especially in EIA1E and EIA1L), although appreciable differences were spotted during the Orientalizing and Archaic periods.
    \item In the case of model L-L, global efficiency $\Eglob$ was almost always lower than that of the empirical networks. Specifically, obtained values presented a high variability across realizations, and very few of them reached values close to the empirical ones. Since, according to this model, nodes had no cognition of the global scale, a good result in terms of global efficiency might occasionally be achieved by chance. On the contrary, the G-L model produced networks whose $\Eglob$ was almost identical to that of their empirical correspondents (usually slightly lower) with extremely limited fluctuations. Networks obtained with model EE also showed similar values of $\Eglob$ and, as opposed to model G-L, they turned out to be slightly more efficient. Both these results were expected because the quantity $R$ is directly related to the global efficiency (see. Eq~\ref{eq:eglob}).
    \item The values of the local efficiency $\Eloc$ in synthetic networks produced by model L-L were significantly higher than those of their empirical counterparts. It was just the opposite for the values obtained from synthetic networks produced by model G-L, which were lower. The closest approximation to $\Eloc$ was yielded once more by the EE model. The agreement with the local efficiencies of reference is particularly precise since EIA2. Only in the first Age, when model G-L and model EE were practically equidistant to the target value, none of the models could provide completely satisfactory results. 
\end{enumerate}
Summarizing, model L-L had the best local connectivity (high $\avg C$ and $\Eloc$), but performed very poorly at the global scale (low $\Eglob$), while model G-L was weaker at the local scale (low $\Eloc$, nonexistent $\avg C$), but quite efficient at the global level. More interestingly, model EE displayed the highest values of global efficiency, while doing reasonably well in terms of local efficiency. Both in model G-L and in model EE, nodes explicitly sought to increase the global efficiency by means of the creation of shortcuts. However, the clever, thrifty use of resources implemented in the EE model, prioritized shorter links. The improvement in the connectivity at the local scale was obtained as a useful byproduct.
\FloatBarrier
                        
\subsection{Network difference}
\begin{figure}
\centering
\includegraphics[width=0.8\columnwidth]{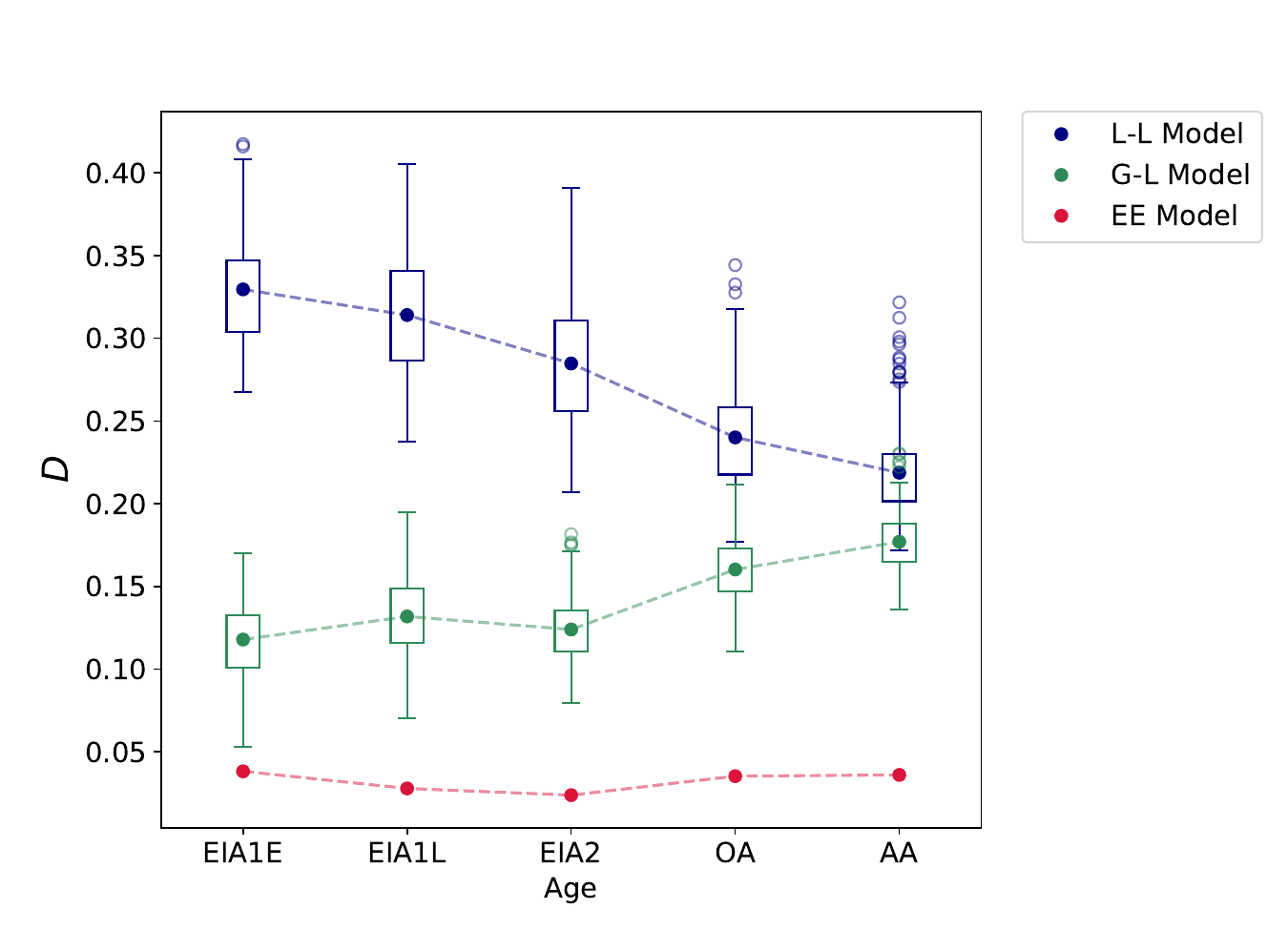}
\caption{Difference $D$ from the empirical to the synthetic networks.}
\label{fig:dist}
\end{figure}

The previous subsection revealed successful and unsuccessful aspects of each model, and seemed to place the EE model and its coordinated decision making ahead of the ones that included a competitive mechanism. However, evaluating the performance of the models by merely looking at the properties one by one is not accurate-enough. To assess the overall agreement between the properties of empirical and synthetic networks (i.e. by considering all four of them at once), we defined the following difference function $D$:
\begin{equation}\label{eq:difference}
D = \frac{1}{4} \left( \frac{|\avg l^e - \avg l^s|}{\avg l^e} + |\avg C^e -\avg C^s| + |\Eglob^e -\Eglob^s| + |\Eloc^e -\Eloc^s|\right)
\end{equation}
where $e$ and $s$ indices stand for empirical and synthetic networks respectively. According to this expression, the difference between two networks is the average of the absolute differences of the four characterization measures. The only measure not defined in the range $[0,1]$, the average link length, was normalized dividing by its value in the empirical system. Figure~\ref{fig:dist} shows the behavior of the three models in terms of the difference $D$ across the Ages.
The deterministic model provided the smallest $D$ for every age. The initial period EIA1E showed the smallest difference between model G-L and model EE. In fact, this was the only moment in which a few runs (around $1\%$) of model G-L produced similar values of the difference function. More in general, we can conclude that, through time, the EE model consistently outperformed the others. 
\FloatBarrier
\section{Discussion}\label{sec:disc}
\subsection{Iron Age southern Etruria. Evaluation of the results.}
Observing how close the behavior of model EE is to that of the empirical networks, our results suggest that this mechanism for the creation of connections resembles, although in an extremely simplified way, the essence of transportation network configuration in Etruria during the Iron Age. Although it is always possible that a different model (implementing different mechanisms) would also recover the features of the empirical road connections, from a general historical viewpoint, our model represents a realistic scenario. It is reasonable that cities and towns from those times were interested in being connected to each other, knew their individual necessities, and shared such information at the regional scale. 

It is also worth noting that the underlying assumption about what the node-polities knew - which is the same in both model EE and model G-L - is also realistic. 
Indeed, models G-L and EE do not assume that cities had a complete knowledge on the state of the whole system at each moment, which is a kind of information that would have hardly been available to them. On the contrary, according to these two models, each one of them only knew its own situation with respect to all the other sites, that is, had a complete subjective map of the territory, and tried to improve its connectivity in the most profitable way.

When considering the details of the EE model, the hypothesis that local interests were sorted according to a more general (global) assessment of the overall functioning of the region is compatible with the existence of a league of loosely cooperating independent city-states. To hypothesize a general common criterion that guided consensus about how resources had to be allocated when dealing with something as fundamental as roads, is perfectly plausible in this scenario. 

Of course, we do not propose that somebody was literally computing the value of the ratio between the length of the new route and the shortest existing path between any pair of places. However, it is reasonable to imagine that this was what polities actually did, even though in an approximate, qualitative way. The translation of the insights obtained from simple formal models into the language of the historical processes is necessarily a loose one. Beyond the mathematical formalism, the idea of comparing costs and benefits in terms of shortening connections is a useful concept. 

Finally, we would like to focus on the political initiatives of towns and cities. It is important to stress that - not only in model G-L, but in the EE model as well - they were never explicitly pursuing the global good (which would not be compatible with their political independence). In both cases, the good quality of the road network at the global scale (global efficiency), is accomplished as an unintended consequence of the node-polities seeking their individual benefit.
Furthermore, the EE model - the one which reproduces the most significant features of an empirical road network - balances the individualistic interests of cities and towns by incorporating some level of cooperation, or at least coordination, among the different polities and consider them as equally important, thus creating even better networks. Such a mechanism agrees with our knowledge about the organization of the Etruscan cities.
\subsection{Inferring inter-polities organization from TTI: an achievable goal.}
We have argued so far that the EE model does not contradict any previous knowledge on the case under study. Taking a step further, we would suggest that the fact that the EE model points to \textit{communication} and \textit{balanced powers} as ingredients that are essential to build road networks, is a necessity and not an accident. Indeed, such ingredients are basic traits of the political organization of the Etruscans, generally defined by scholars as an heterarchical organization (see recently \citealp{stoddart_2017} as opposed to the more structured and hierarchical organization of the Roman polity \citep{fulminante_and_stoddart_2012}. In conclusion, we are quite confident that a good mechanistic model for TTI at the regional scale reflects and captures fundamental aspects of the relationships between polities.

Consequently, if a good model for road networks reflects the nature of the interactions between sites, then it is possible to infer basic aspects of such interactions when they are unknown by means of the formal study of the routes that connected them. To strengthen such a conclusion, it would be useful to compare the Etruscan case with other benchmark case-studies for whom we know both routes and political organization (see for example the case study of the Inca Empire, for which see \citealp{jenkins_2001}, or hollow routes of 3rd Millennium BC Mesopotamia analyzed by \citealp{menze_2012}).
\subsection{Population, territory, and agency. Limits of the applicability of the methodology.}

So far, we have presented our approach as a methodology for studying the organization of a system of settlements at the regional scale. Beyond the generic concept of ``regional scale'', there exist some relevant features that are pre-requisites for the applicability of the proposed analytical method, as it has been devised in the present work. Such pre-requisites concern three fundamental aspects of the system under study: the geographical distribution of the population, the agency of the settlements, and the geometric shape of the territory.

Regarding the population, human communities have to be localized in a number of geographically separated places, \ie cities and towns, rather than being fragmented into small villages and farms, or continuously distributed over vast urban areas. In other words, it is essential that the geographic distribution of the population is suitable to be approximated as a set of geolocalized nodes. Only under this condition, we can regard roads as connections linking places, instead of mere paths through them. Additionally, the number of such places has to be large enough for the network metrics to be used as meaningful indicators. Usually, few tens is enough for this purpose.

However, the geographic localization of past communities is not a sufficient condition. They need to have agency over the matter of building roads and therefore, depending on the circumstances, it may be appropriate to filter out smaller sites. On the other hand, the presence of cities surrounded by dependent towns is not an obstacle whenever it is reasonable to assume -- as we did for southern Etruria --  that the interests of the cities are split into more than one location.
At the systemic level, some caution should be taken when applying the proposed approach to non self-ruled regions whose TTI are part of a larger scale infrastructure. Similarly, territories made of a large number of sub-units (\eg Empires and provinces) are also problematic since the (hypothetical) consensus for building local road is hardly reached at the global scale. If a case study possesses such characteristics, we should consider the possibility of modifying the hypotheses about the decision-making mechanisms proposed in this work and the corresponding network models. For instance, it could be very interesting to include, besides the local and the global scale, an intermediate scale (mesoscale) at which the consensus could be reach without involving the whole system. Furthermore, the hypothesis of a centralized authority planning the design of the TTI -- here discarded -- may be sensible for other political contexts.

Finally, the shape of the region also matters. We should be cautious when applying models based on geodesic distance in highly non-convex region where a large fraction of potential links crosses its borders or physical boundaries. Such links have to be regarded as forbidden connections. In southern Etruria there were some lakes and a number of promontories, but since they were small, we could neglect them. On the contrary, if we would apply our methodology to regions such as, for instance, the whole Italian Peninsula, this issue should be addressed.

Summarizing, the extension of the area under study is not an issue. The methodology can be applied to small or large regions, whenever their population was localized in a number of separated places and they had political authority over the construction of the ways that connected them. However, if their shape was highly irregular, this aspect needs to be included in the modelling.

\subsection{Beyond straight lines and perfect equality. Possible directions for improvement.}
One of the most valuable characteristics of our models is their elegance and simplicity. They produce easily distinguishable results that can be interpreted in a clear way. Nonetheless, it is clear that the extreme lack of sophistication of our approach has a price.

Adopting the geodesic distance as the estimate for link cost is likely the most significant approximation. Unfortunately, there is no gradual way to improve the precision of this measurement. As we already discussed in Section \ref{sec:map2net}, to implement GIS based techniques comes with serious implications for the whole modelling process. However, it can be useful to analyze what could be done in the hypothetical case of having the complete matrix of least cost paths between every pair of nodes. Above all, it is crucial to account for the presence of intersections between paths, something that we can disregard easily in our abstract framework, but which are an important element of real road maps. Thanks to crossroads, two nodes can be connected by means of direct paths that do not pass through any other node, but that are not the LCP between them either. At the same time, some stretches (\eg those between two intersections) are part of more than one LCP. Once a path including a stretch of this kind is built, the cost of all the paths that include it is reduced. As a consequence, we are no longer allowed to identify the cost of connecting two nodes with the length of the connection between them. Therefore, Eq~\ref{eq:detour} is not a good proxy for benefit-cost ratio.

In conclusion, the complexity of a modelling approach based on LCP would be much higher than that of the baseline models presented in this work. Our purpose was setting the ground for a new quantitative methodology to tackle regional TTI and including LCP goes beyond the scope of this stage of the research. Nevertheless, it is undoubtedly a promising direction for future investigation.

Additionally, future efforts have to be devoted to the generalization of the model presented in this work. We have been lucky enough to start with a case-study that could be very well fitted to a simple model. We can assume that the basic mechanisms are present in other similar case-studies, but not necessary with the same intensity. In a different scenario, coordination among political entities could be less perfect and a certain degree of stochasticity would be required. Moreover, the complete equality that worked so well for Etruria might need to be relaxed for other cases. 

A good example of this is Iron Age Latium vetus, a region neighboring southern Etruria which we addressed in a recent study \citep{FPML2017}. To attain a satisfactory reproduction of the main features of empirical road network, we had to modify model EE by including a tunable amount of rich-get-richer bias.
In any case, since EE model is deterministic and has no parameter, it is readily suited for modification to easily capture these and other particularities.

\subsection{Other uses of the EE model: inverse engineering to infer unknown road networks}
Model EE has been devised to infer information about the organization of a system of settlements using TTI as an indirect proxy. Even though further investigations is needed, it has so far proven to be a good model for transportation networks in contexts of local interests balanced through coordination mechanisms. We will now consider the opposite scenario: a known type of organization, compatible with the EE model, and an incomplete road map.

In this case, if it was possible to formulate alternative hypotheses about the unknown part of the TTI under study, which could be translated into alternative empirical networks, it would be then feasible to use model EE to help recovering the missing information. The basic idea is that the empirical network with the highest similarity with its model generated counterpart is the most plausible. In other words, we can reverse engineer the tool, selecting the best empirical network instead of the best artificial one. This approach can be applied to infer either the position of one or more nodes or the existence of some connections. On the contrary, if it is not possible to come up with alternative hypotheses, the connectivity proposed by the EE model could still be assessed qualitatively\footnote{The model has already proved useful to better asses some parts of the network that were more hypothetical, such as those paths suggested by the position of the settlements}. In future work we are planning to refine this and other modelling techniques (also within a GIS environment, in collaboration with German colleagues, \citealp{faupel_2018} and \citealp{faupel_and_nakoinz_2018}) to produce high resolution maps of pre-Roman paths and compare them with Roman roads to assess continuity and/or discontinuity between the two systems.

From a practical point of view, since the total link length cannot be defined, the stopping condition is one of the most critical aspect. If the unknown roads are all and only those connecting a certain set of nodes, then the algorithm stops adding new links when the total length of the connections between nodes whose connectivity is known is the same as in the empirical network\footnote{Major roads connecting primary settlements in the Roman provinces of Hispania (218 BC - 5th century) are an example of this situation. We are currently working on this case-study in collaboration with Pau de Soto and Tom Brughmans.}. For other more complex possibilities, specific solutions should be devised, while in the extreme event of a completely unknown road map, the total amount of connection length has to be determined by means of heuristic arguments. From a different perspective, in their recent work, Groenhuijzen and Verhagen (\citealp{Groenhuijzen_and_Verhagen_2017}) applied several network models -- among them, a slightly modified version of our EE model -- to the case of a TTI about which nothing was known, neither the underlying mechanisms, nor any of the paths that were part of it. The only available information was the purpose of the TTI and the authors proposed to exploit this knowledge to determine which model produces the most plausible networks.

Hence the number of possible applications in this direction is vast. For instance, the EE model provides a different algorithm to build geographical networks that could complement the usual nearest neighbors rule adopted for proximal point analysis.

\section{Conclusions}\label{sec:conc}
Roads and paths have been the object of study of antiquarian, landscape and topographical research since the eighteenth and nineteenth centuries. With reference to Etruria, several studies have emphasized the importance of road building and TTI creation for the development of cities and their control over the countryside, especially in the Orientalizing and Archaic Periods. However, these studies mainly adopted a qualitative perspective. 

By applying newly developed Network modelling technique to a well-known case-study such as Etruria, this paper demonstrated the importance of studying TTI not only with a qualitative approach but also from a quantitative perspective, and has identified opportunities for further developments in numerous directions.

Firstly, a quantitative approach allows the study of long-term perspectives and/or the comparison between different polities in the same region and/or different geographical and chronological contexts allowing further cross-cultural comparisons of regional systems transportation networks and peer-polities interaction;
Secondly, the application of modelling techniques to a known historical case-study has allowed the development of a basic toolbox that can be readily modified for addressing potentially any kind of scenario, thus allowing for the creation of a novel methodology for shedding light on the nature of the interactions between past human communities.
\section*{Acknowledgments}
LP and IM are supported by the European Research Council Advanced Grant EPNet (340828). FF was supported by the Marie Curie programme, FP7-PEOPLE-2013-IEF, Past-people-nets (628818). She is also thankful to the Institute of Advanced Study (IAS) at Durham which has provided fellowship and a favourable research environment for the revision and completion of the paper. SL acknowledges  financial  support  from  the Ramón y Cajal programme through the grant RYC-2012-01043 and the Generalitat de Catalunya (Project No.2017SGR-836).
\bibliography{biblio}
\end{document}